\documentclass{ws-procs9x6-cpt16}

\begin{document}

\newcommand{\refeq}[1]{(\ref{#1})}
\def\etal {{\it et al.}}

\title{Classical Nonminimal Lagrangians and \\ 
Kinematic Tests of Special Relativity}

\author{M.\ Schreck}

\address{Departamento de F\'{i}sica, Universidade Federal do Maranh\~{a}o\\
S\~{a}o Lu\'{i}s, Maranh\~{a}o 65080-805, Brazil}

\begin{abstract}
This article gives a brief summary on recently obtained classical lagrangians for the nonminimal fermion sector of the Standard-Model Extension
(SME). Such lagrangians are adequate descriptions of classical particles that are subject to a Lorentz-violating background field based on the SME.
Explicitly, lagrangians were obtained for the leading nonminimal contributions of the $m$, $a$, $c$, $e$, and $f$ coefficients. These results
were then used to interpret classical, kinematic tests of Special Relativity in the framework of the nonminimal SME. This led to new constraints
on certain nonminimal controlling coefficients. Although the experiments were very sophisticated in the era when they were carried out, their
sensitivities for detecting Lorentz violation were still far away from the Planck scale. Obtaining the novel constraints can be considered as a
proof-of-principle demonstrating the applicability of the classical lagrangians computed.
\end{abstract}

\bodymatter

\section{Introduction}

The Standard-Model Extension (SME) is an effective field-theory framework parametrizing possible deviations from Lorentz invariance in the Standard Model\cite{Colladay:1996iz,Colladay:1998fq} and General Relativity.\cite{Kostelecky:2003fs} Its formulation is model independent and its realm of
applicability ranges from the very low energies of atomic hyperfine splitting to the ultra-high energies of certain cosmic rays, which still
lie much below the Planck energy, though. The SME neither modifies the gauge structure of the Standard Model nor does it introduce new particles.
In contrast, it modifies some of the particle properties that are connected to Lorentz invariance. This concerns dispersion relations, field
equations, their solutions,~etc. The SME can be considered as an expansion in terms of derivatives. Its minimal version involves all component
coefficients that are contracted with field operators of mass dimension three or four whereas the nonminimal SME contains an infinite number of
component coefficients contracted with higher-dimensional operators. The expansion is furnished such that the number of derivatives in the
field operators successively increases by two, as does the number of indices of the component coefficients.\cite{Kostelecky:2013rta}

\section{Classical lagrangians}

The SME is a field-theory framework, which is why it is reasonably applied to describe quantum processes of elementary particles in colliders or
for ultra-high energy cosmic rays. In contrast, when describing the motion of a macroscopic test body in the gravitational field of the Earth
it is much more reasonable to work within a classical framework based on the SME. Therefore, it is desirable to construct a map from the field
theory description provided by the SME and based on a Lagrange density to the Lagrange function of a classical, relativistic, pointlike particle
moving with four-velocity $u^{\mu}$. The map is provided by a set of five nonlinear ordinary equations that involve the momentum $p_{\mu}$ of
the field-theory description, the four-velocity $u^{\mu}$ of the classical description, and last but not least the classical lagrangian $L$.
These equations read\cite{Kostelecky:2010hs}
\begin{equation}
{\cal R}(p)=0\,,\quad \frac{\partial p_0}{\partial p_i}=-\frac{u^i}{u^0} \text{ for } i\in \{1,2,3\}\,,\quad L=-p_{\mu}u^{\mu}\,.
\end{equation}
The first is the dispersion equation of the particular Lorentz-violating fermion sector under consideration. Besides the four-momentum
components, it involves Lorentz-violating controlling coefficients. The centroid of a wave packet is supposed to propagate with the group
velocity assigned to it. Since such a wave packet is localized in space its classical limit can be considered as a particle.
Hence, the second, third, and fourth equations link the group velocity of a wave packet in the field theory to the three-velocity of the
classical particle. The minus sign on the right-hand side takes into account the different position of the spatial index $i$ on both sides.
Finally, the last equation involves the lagrangian. It follows from the condition of positive homogeneity, $L(\lambda u)=\lambda L(u)$
for $\lambda>0$, which guarantees that the corresponding action does not depend on the parametrization of the particle trajectory.
These equations must be solved for the lagrangian, which should be expressed in terms of the four-velocity. This was carried out for
various sectors of the minimal SME at all orders in Lorentz violation. An investigation performed lately even shows that it is possible
to obtain a classical lagrangian for an arbitrary case of the minimal fermion sector at first order in Lorentz violation and at second
order in the particle velocity.\cite{Schreck:2016jqn}

\subsection{Lagrangians for nonminimal coefficients}

A recent goal was to extend the set of known classical lagrangians to at least some feasible cases of the nonminimal SME. Since the structure
of these lagrangians was supposed to be highly involved (cf.\ Ref.\ \refcite{Schreck:2014hga} for the exact result for a nonzero nonminimal coefficient
$m^{(5)}_{00}$) the analysis was restricted to first order in Lorentz violation. It was practical to solve the set of nonlinear equations
using the method of Gr\"{o}bner bases. Obtaining such a basis for the set of nonlinear polynomials is comparable to bringing a system of linear
equations into triangular form. Thus, a Gr\"{o}bner basis at hand allows for a convenient solution of the nonlinear system.

Classical lagrangians were obtained for the lowest-dimensional nonminimal terms in the momentum expansion of the SME for the $m$, $a$, $c$, $e$,
and $f$ coefficients. The corresponding field operators for the $m$ and $a$ coefficients are of mass dimension five whereas the field operators
for the $c$, $e$, and $f$ coefficients have mass dimension six. For reasons of comparison, both the minimal and the nonminimal results
corresponding to particles are stated for the $a$ coefficients at first order in Lorentz violation:\cite{Kostelecky:2010hs,Schreck:2015seb}
\begin{eqnarray}
L^{\widehat{a}^{(3)\mu}}&=&-m_{\psi}\sqrt{u^2}-\widehat{a}^{(3)}_{\ast}\,,\quad \widehat{a}_{\ast}^{(3)}\equiv a_{\mu}^{(3)}u^{\mu}\,, 
\nonumber\\
L^{\widehat{a}^{(5)\mu}}&=&-m_{\psi}\sqrt{u^2}-\frac{m_{\psi}^2\widehat{a}^{(5)}_{\ast}}{u^2}+\dots\,,\quad \widehat{a}^{(5)}_{\ast}\equiv a^{(5)}_{\mu\nu\varrho}u^{\mu}u^{\nu}u^{\varrho}\,,
\end{eqnarray}
where $m_{\psi}$ is the particle mass. Similar results were obtained for the $m$, $c$, $e$, and $f$ coefficients. The latter lagrangians are
perturbative, which is why they are sums of the standard result $L=-m_{\psi}\sqrt{u^2}$ and a Lorentz-violating contribution. In general, such
a lagrangian is not unique. However, by dimensional reasons, its shape is very restricted when restrained to first order in Lorentz violation.
Due to observer Lorentz invariance, the Lorentz-violating term can only involve contractions of the component coefficients with the
four-velocities as these are the only tensor quantities that can form observer Lorentz scalars. Furthermore, since the mass dimension of the
component coefficients $a^{(5)}_{\mu\nu\varrho}$ is lower by two compared to the minimal ones the new term must include an appropriate power
of the particle mass to make the mass dimension of this contribution consistent with the dimensionality of the standard term. Last but not
least, because of the additional powers of the four-velocity contracted with the component coefficients, the denominator has to contain a
suitable Lorentz scalar formed from the four-velocity. After all, the dimension of velocity in the Lorentz-violating summand should match the
dimension of the standard~result.

\section{Experimental constraints}

Finally, two experiments testing the kinematics of Special Relativity were considered. Since these tests are purely kinematical their results
were evaluated using classical lagrangians. In the first experiment, which was table-top, an electron was arranged to travel through a homogeneous
magnetic field where it was deflected to traverse a circular electric field afterwards. Such an arrangement allowed for measuring the particle
momentum and its velocity-dependent mass. Since it was challenging to measure the electric field to a sufficient precision at the time when the
experiment was performed the experiment was repeated with protons. With the subsequently obtained values for the proton momentum and mass,
the electric field could be eliminated from the equations. Under the assumption that a Lorentz-violating signal hides within the double of the
average experimental error, bounds on certain nonminimal coefficients were computed. The sensitivity of the experiment was shown to be far away
from the Planck scale.\cite{Schreck:2015seb}

The second test considered was an accelerator experiment performed at SLAC. Electrons were accelerated to a certain energy and arranged to hit a
thin target to produce \textit{bremsstrahlung}. Behind the target both the scattered electron and the \textit{bremsstrahlung} photon traveled a
long distance until they were converted to positrons. With an rf separator the positron originating from the electron was spatially separated from
the positron produced from the photon. The spatial separation was proportional to the arrival time difference of the electron and photon, i.e.,
this setup measured velocity differences between the initial particles. This allowed for obtaining another set of constraints on the nonminimal
electron sector. The higher Lorentz factor increased sensitivity, which led to an improved set of bounds.\cite{Schreck:2015seb}

\section*{Acknowledgments}

The author acknowledges travel support by FAPEMA.

\end{document}